\documentclass[sigplan,screen,sigconf]{acmart}

\usepackage{geometry}
\usepackage{graphicx} 
\usepackage[english]{babel}
\usepackage{blindtext}
\usepackage{multicol}
\usepackage{todonotes}
\usepackage{lipsum}  
\usepackage[absolute,overlay]{textpos}

\setcopyright{rightsretained}
\acmDOI{}
\acmISBN{}
\acmConference[LATTE '21]{1st Workshop on Languages, Tools, and Techniques for Accelerator Design}{April 15, 2021}{Virtual, Earth}

\title{Enabling Cross-Domain Communication: How to Bridge the Gap between AI and HW Engineers}
\author{Michael J. Klaiber, Axel J. Acosta, Ingo Feldner, Falk Rehm}
\email{firstname.surname@de.bosch.com}
\orcid{1234-5678-9012}
\affiliation{
  \institution{Robert Bosch Corporate Research, Renningen}
  \streetaddress{Robert Bosch Campus 1, Renningen, Germany}
  \country{Germany}
}

\renewcommand{\shorttitle}{Bridge the Gap between AI and HW Engineers}
\begin{document}

\begin{abstract}

  A key issue in system design is the lack of communication between hardware, software and domain expert. Recent research work shows progress in automatic HW/SW co-design flows of neural accelerators that seems to make this kind of communication obsolete. Most real-world systems, however, are a composition of multiple processing units, communication networks and memories. A HW/SW co-design process of (reconfigurable) neural accelerators, therefore, is an important sub-problem towards a common co-design methodology. The ultimate challenge is to define the constraints for the design space exploration on system level - a task which requires deep knowledge and understanding of hardware architectures, mapping of workloads onto hardware and the application domain, e.g. artificial intelligence.
  For most projects, these skills are distributed among several people or even different teams
  which is one of the major reasons why there is no established end-to-end development methodology for digital systems. This position paper discusses possibilities how to establish such a methodology for systems that include (reconfigurable) dedicated accelerators and outlines the central role that languages and tools play in the process.

\end{abstract}

\maketitle

\section{Introduction}

Effective hardware/software co-design methodologies have been examined for decades and are still not the norm
in the development of large-scale (embedded) systems \cite{slomka2000,lima2015}.
This often results from sequential or circular dependencies in the concept or requirements phase of a project.
In particular, this means 
that hardware architects need workload specifications to dimension communication and computational resources, and application domain engineers, such as AI algorithm engineers, require information about the hardware before they can work efficiently.
As this paper deals mostly with AI workloads, in particular neural networks, we use the term AI engineer
synonymously with application domain engineer or software engineer to describe the person who transforms a specification or solution to a problem
into executable code.

Many development methods such as the waterfall model,
however, do not account for ramifications of algorithm to hardware and vice versa.
Even if algorithm developers, software developers or hardware developers use agile development methods,
there is no established communication medium to exchange requirements between their development paths.
Assumptions in the different disciplines can't be exchanged easily.
A result is that hardware is designed with assumptions
about a workload which might already be outdated, and the algorithm is designed without an understanding what
operations provide good performance on the target hardware.

This holds especially true in the field of embedded systems, as these systems often have low computing and communication resources
and heterogenous project specific platforms.

The methodology proposed in this paper should help to answer questions like:
\begin{itemize}
  \item What implication does a specific algorithm change have on the system performance?
  \item What are the operations that are worth to be accelerated in hardware?
  \item Would a rearranging of physical memories or memory layout improve performance?
  \item How much more chip area and/or power do I want to spend for x\% more accuracy or y\% faster execution time?
  \item How to optimize an algorithm to perform best on a target platform without any hardware changes?
\end{itemize}

State-of-the-art methods show sophisticated methods for
\begin{itemize}
  \item HW/SW co-design and generation of neural network accelerators \cite{hao2019, shi2020},
  \item Virtual Hardware Models \cite{klaiber2019} and 
  \item Generation of Virtual Models from RTL Code \cite{snyder2017}. 
\end{itemize}
The ideas presented in the following are mostly possible due to the recent progress in neural network compiler stacks, such as TVM \cite{chen2018} or Glow \cite{rotem2018}, used in combination with
 with virtual hardware and system models as bridging technology.

\section{Methodology}

\begin{figure*}
  \begin{center}
  \vspace{-5mm}
  \begin{multicols}{2}
      \includegraphics[width=\textwidth]{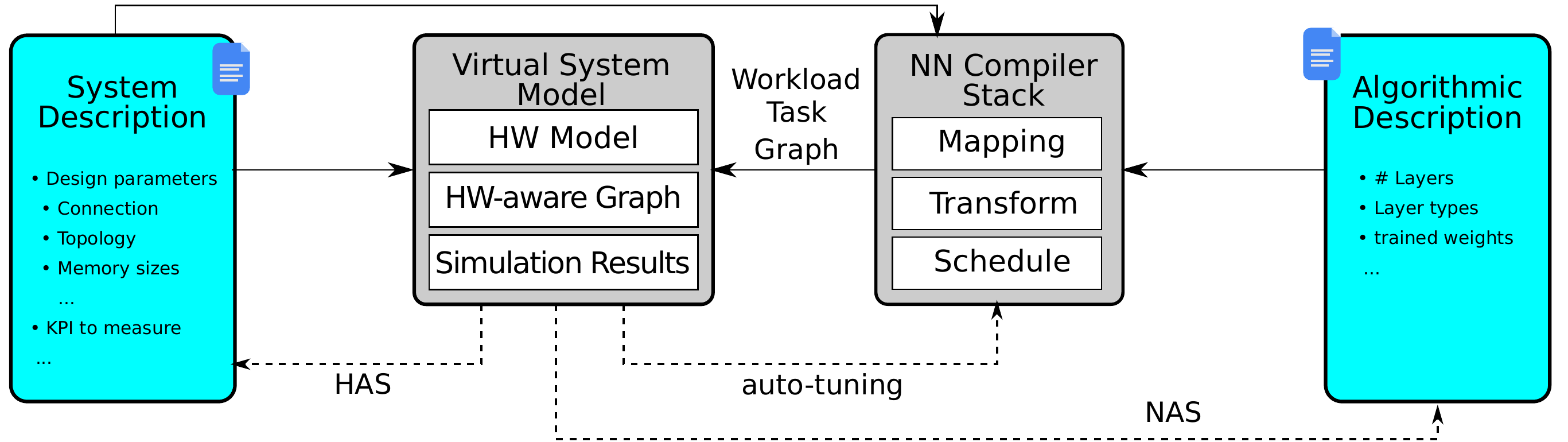}
  \end{multicols}
  \vspace{-7mm}
  \caption{Development flow to enable system level cross-domain communication.
    \label{fig:methodology}}
  \vspace{-2mm}
  \end{center}
\end{figure*}
  
  \subsection{The AI Engineer's Perspective}
  The primary objective of most AI Engineers is to design algorithmic models that achieve high prediction accuracy for a given data set. This model defines the workload to be executed on the target hardware. 

  In Fig. \ref{fig:methodology}, the AI engineer's perspective is represented by:
\begin{itemize}
 \item the \textbf{Algorithmic description}. This is an abstraction layer created by AI practitioners to formalize a problem with regards to algorithmic properties. Common formats are TensorFlow and ONNX.
\end{itemize}

The dominating technology that AI Engineers use for neural network training are GPUs and most likely server grade infrastructure. The execution time and memory footprint are, therefore, often secondary metrics that are evaluated based on the GPU's architectural properties and toolchains. This represent the first break of synergy of a AI/SW/HW team striving for joint development.
An established workaround, is the use of cost functions to estimate low level implementations\cite{elsken2019}. 
The difficulty arises in defining sophisticated and accurate cost functions for the target hardware. A task mostly handed over to the next stakeholders.

  \subsection{The Hardware Architect's Perspective}

  Hardware architects have many different tasks, among which particular important ones are to identify workload patterns that can be accelerated by dedicated computation blocks and to change the physical structure of the system to improve performance.

  Workload patterns that occur often in a system can be accelerated by the use of special instructions on a CPU or by dedicated accelerators.
  Both possibilities require a detailed and formalized description of the workload to be executed. Whenever this
  workload
  description changes, considerations about the dedicated computation blocks can become obsolete.

  The physical connections, between computation blocks, communication networks and memories play a major role how fast a workload can be
  executed. Again, here a formalized workload description is the basis for decisions, e.g. to determine the size and granularity
  of the memories,  how much bandwidth to allocate between a computation block and a memory, etc.
  A change of the workload description can make these considerations obsolete or inefficient.
  
 In Figure \ref{fig:methodology}, the HW architect's perspective is represented by:
\begin{itemize}
  \item The \textbf{System description}: is an abstraction layer that provides a formalized description provided by the hardware
  architect. It contains the physical connections of hardware components, timing information of each hardware component requires
  to process an intrinsics value, and transaction behavior. For instance the time required to perform Conv2D instrinsic on an
  accelerator or the time to transfer 1 kByte of data from internal memory to the accelerator while five other hardware components
  request bus access. A common format in the automotive system context is Amalthea \cite{amalthea17}.

  \item The \textbf{Virtual hardware model}: is the executable form of the system description interfacing with the compiler.
  It essentially transforms the workload call graph into simulated execution time, memory accesses or other KPI.
\end{itemize}  

HW development is a high effort and costly process. Therefore, there are few opportunities to make wide-reaching changes or adaptations after an architecture has been frozen. In order to cope with changes in the workload (or application) and enabling easier programmability, all computing HW systems must also offer a SW stack to automate the generation of implementations.

\subsection{The Compiler Developer's Perspective}

The compiler is the SW tool which bridges the abstraction between the AI engineer's domain and the HW architect's domain. It is given \emph{the} daunting task of exploiting all HW properties designed into the architecture in order to create a good (or best) implementation of the AI engineer's model. 
 
 In Fig. \ref{fig:methodology}, the compiler developer's perspective is represented by:
\begin{itemize}  
\item The \textbf{Neural Network compiler stack}: The implementation of the model onto the hardware requires a mapping from AI engineer's domain to intrinsics of the (virtual) target hardware. For the mapping process it requires the intrinsics of all hardware components
  in the system. For optimization, the compiler needs timing and resource usage information of the hardware components. All of this information is contained
  in the system description. The transformation and optimization steps essentially formalizes a set of rules and knowledge
  that both hardware architect
  and AI engineers have about the system and the workload, and therefore, build the possibility for a fully automated exchange of information between the hardware architect and AI engineers.

  \item The \textbf{Workload task graph} is an abstraction layer consisting of a directed graph where each node represents an intrinsic
  call  mapped to a hardware component of the system description. Each edge represents dependency relations. The task graph, for
  instances, describes a dependency like: the DMA transfer from extern memory to accelerator memory must be finished before computation on
  the transported data can be started.
 \end{itemize} 

\subsection{From Isolated Optimizations to Holistic Solving}

  The dashed lines in Figure \ref{fig:methodology} show the paths for (automatic) optimization based on key performance indicators (KPI).

In isolation, each of the perspectives of the previous subsection has its own optimization problem and solutions:
\begin{itemize}
\item The AI engineer: optimizes the accuracy of a model via Network Architecture Search (NAS) methods\cite{elsken2019}.
\item The HW architect: optimizes the HW resources via Hardware Architecture Search (HAS) methods \cite{jiang2020}.
\item The Compiler developer: optimizes the implementation via auto-tuning methods\cite{chen2018}
\end{itemize}

The methodology we sketch in this paper tries to fill the gap that results from a lack of human communication in medium to large teams.
However, the long-term goal of the community should be to extend this idea to combine all the optimization problems into a single one. 

The conceptual break between these three optimization domains, would lend itself to a natural division of labor which leads to domain silos driving 3 different tool development.
This, in our view, is a considerable organizational hurdle which will prevent effective communication between the teams.
 A solution to this, we argue, is to see only the one problem: "Creating a complete HW/SW stack for your product".
Such a holistic view should also come with a more heterogeneous organization of teams, for example have a number of HW experts working on the higher levels of the stack (and vice versa).
It is in such a setting that HW/SW/System co-design will become mainstream in large embedded system design.

\section{Conclusion}
HW/SW co-design for (reconfigurable) accelerators needs to take other components of the system into account. Compilers and languages are the key element to create a methodology that bridges the gap between AI and hardware engineers. In this paper we sketched a possible methodology and its components based on (mostly) existing technologies and show a new way how to bring optimization to the system level.

\bibliography{latte}

\end{document}